# Tuning the electronic properties of hydrogen passivated $C_3N$ nanoribbons through van der Waals stacking

Jia Liu[1#], Xian Liao[1#], Jiayu Liang[1], Mingchao Wang,[2] Qinghong Yuan[1,3*]

[1] State Key Laboratory of Precision Spectroscopy, School of Physics and Electronic Science, East China Normal University, 500 Dongchuan Road, Shanghai 200241, China
[2] Department of Materials Science and Engineering, Monash University, Clayton, VIC 3800, Australia
[3] Centre for Theoretical and Computational Molecular Science, Australian Institute for Bioengineering and Nanotechnology, The University of Queensland, Brisbane, QLD, 4072, Australia

The two-dimensional (2D) $C_3N$ has emerged as a material with promising applications in high performance device owing to its intrinsic bandgap and tunable electronic properties. Although there are several reports about the bandgap tuning of $C_3N$ via stacking or forming nanoribbon, bandgap modulation of bilayer $C_3N$ nanoribbons ($C_3NNRs$) with various edge structures is still far from well understood. Here, based on extensive first-principles calculations, we demonstrated the effective bandgap engineering of $C_3N$ by cutting it into hydrogen passivated $C_3NNRs$ and stacking them into bilayer heterostructures. It was found that armchair (AC) $C_3NNRs$ with three types of edge structures are all semiconductors, while only zigzag (ZZ) $C_3NNRs$ with edges composed of both C and N atoms (ZZ-CN/CN) are semiconductors. The bandgaps of all semiconducting $C_3NNRs$ are larger than that of $C_3N$ nanosheet. More interestingly, AC-$C_3NNRs$ with CN/CN edges (AC-CN/CN) possess direct bandgap while ZZ-CN/CN have indirect bandgap. Compared with the monolayer $C_3NNR$, the bandgaps of bilayer $C_3NNRs$ can be greatly modulated via different stacking orders and edge structures, varying from 0.43 eV for ZZ-CN/CN with AB'-stacking to 0.04 eV for AC-CN/CN with AA-stacking. Particularly, transition from direct to indirect bandgap was observed in the bilayer AC-CN/CN heterostructure with AA'-stacking, and the indirect-to-direct transition was found in the bilayer ZZ-CN/CN with AB-stacking. This work provides insights into the effective bandgap engineering of $C_3N$ and offers a new opportunity for its applications in nano-electronics and optoelectronic devices.

# 1. Introduction

Two-dimensional (2D) van-der-Waals (vdW) crystals, such as graphene, transition metal dichalcogenides (TMDs), and black phosphorus, have recently emerged as a class of novel materials, with their 2D nature offering unprecedented opportunities for their applications in nanoscale devices[1-3]. For example, graphene is regarded as a promising candidate for flexible electronics due to its excellent electronic[1], thermal and mechanical properties[4,5]. However, the lack of an intrinsic bandgap in graphene restricts its use as a field-effect transistor (FET) and consequently its applications in electronics [6-8]. Nitrogen doping of graphene serves as an effective approach to open the bandgap of graphene and transform graphene into an n-type semiconductor [9]. Recently, a new 2D hole-free polyaniline ($C_3N$) has been successfully fabricated by polymerization of 2,3-diaminophenazine [10,11]. This material possesses much higher on/off current ratios than graphene[11,12], and has shown attractive physical properties [13,14] for electronic applications. Computational studies have shown that the 2D $C_3N$ is a semiconductor with an indirect bandgap of ~1 eV (HSE level) [13,15].

Engineering the electronic properties of $C_3N$ in a well-controlled manner is crucial for its practical applications in nanodevices. Several methods have been widely used to tune the electronic properties of $C_3N$, including the application of strains or external electric fields[16,17], quantum confinement (nanoribbons) [18-22], and the formation of vdW heterostructures through stacking [23,24]. Bafekry *et al.* found that the electronic properties of few-layer $C_3N$ vary with different stacking orders and layer numbers[25]. The bandgap of multilayer $C_3N$ monotonically reduces with the increase of the layer number, and the application of an external electric field to bilayer $C_3N$ nanosheet leads to the bandgap decrease and a semiconductor-to-metal transition. The nanoribbon counterpart of $C_3N$ offers more tunability in electronic properties because of the unique quantum confinement and edge effect. Several studies have shown that the band feature and gap size can be modulated by the edge termination and ribbon width[21,25,26]. What's more, the hydrogen (H) passivated $C_3N$ nanoribbons are particularly concerned because of their stability. For example, Li *et al*. reported that the H-passivated zigzag $C_3N$ nanoribbons with edges composed of both C and N atoms (ZZ-CN/CN) are semiconductors, and their bandgaps decrease with the enlargement of ribbon width[21]. Bafekry *et al*. studied the bandgaps of $C_3N$ nanoribbons with various types of edge terminations, and demonstrated that the bandgaps of armchair and zigzag $C_3N$ nanoribbons converge to different values with the increase of the width, neither of which are that of the $C_3N$ nanosheet, due to the fact that the conduction band minimum (CBM) or valence band maximum (VBM) are determined by the edge states[25]. Despite of the reports about electronic properties' modulation of $C_3N$ nanosheet via stacking and forming $C_3NNRs$, tuning the electronic properties of $C_3N$ through vdW stacking of H-passivated $C_3NNRs$ has not been addressed.

In this research, we studied the electronic properties of armchair (AC) and zigzag (ZZ) $C_3NNRs$ with three types of edge terminations including CN/CN, CN/CC and CC/CC. It was found that the significant modulation of the electronic properties of 2D $C_3N$ was achieved via cutting into nanoribbons and stacking into heterostructures. Our theoretical calculations demonstrated that all AC-$C_3NNRs$ are semiconductors, in which AC-CN/CN has a direct bandgap while AC-CN/CC and AC-CC/CC have indirect bandgaps. As for ZZ-$C_3NNRs$, only ZZ-CN/CN possesses an indirect bandgap, and ZZ-CC/CN and ZZ-CC/CC are metallic. Particularly, the bandgap modulations of AC-$C_3NNRs$ and ZZ-$C_3NNRs$ with the most stable edge of CN/CN edge are studied through stacking. Our results indicated that the bandgaps of bilayer $C_3NNRs$ with different stacking orders have different reduction in comparison with corresponding monolayer $C_3NNRs$. We also analyzed the energy shift of electronic bands of bilayer heterostructures, and related it to their corresponding orbital overlaps. It was shown that AA- and AB'-stackings have the most and least numbers of orbital overlaps, which is consistent with the values of their energy shift.

## 2. Computational methods

The first-principles density functional theory (DFT) calculations of total energies and electronic structures were carried out by using a planewave basis set and pseudopotentials for describing core and valence electrons[27, 28], as implemented in the Vienna ab initio simulation package (VASP)[29]. Electron exchange and correlation were included through the generalized gradient approximation (GGA) in the PBE form[30]. The hybrid density functional (HSE06) [31-33] was also used to calculate the bandgap in a more accurate way. Spin polarization was considered with a planewave energy cutoff of 500 eV. All structures were geometry-optimized until energy and force differences were converged to $10^{-5}$eV and 0.01 eV/Å, respectively. The van der Waals (vdW) correction proposed by Grimme (DFT-D2) [34-36] was utilized to include long-range vdW interactions for bilayer $C_3N$ nanoribbons. For AC-$C_3$NNRs, the Brillouin zone was sampled using 3×1×1 and 6×1×1 k-meshes for the structural optimization and electronic structure calculations. As for ZZ-$C_3$NNRs, 1×6×1 k-meshes are used to calculate the structural optimization and electronic structure. A vacuum space of 15 Å was used to avoid the neighboring interactions.

## 3. Result and discussions

The unit cell of $C_3N$ structure is composed of six carbon (C) and two nitrogen (N) atoms, which is a flat honeycomb sheet akin to graphene. The calculated lattice constant of $C_3N$ is 4.861 Å, which agrees well with the experimental result of 4.75 Å[10]. Based on the different cutting directions (**Fig. S1** in **Supplementary Information (SI)**) of the $C_3N$ nanosheet, two types of $C_3N$ nanoribbons with different edge configurations are denoted as AC- and ZZ-$C_3$NNRs, respectively. All the $C_3$NNRs' edges are passivated by H atoms to ensure the stability of the ribbon. Different from graphene, the edge structure of $C_3N$ has different compositions due to the mixed components of C and N atoms. As shown in **Fig. 1(a)**, AC-$C_3$NNRs have three combinations of edge configurations, in which two of them have the same edge configuration at both edges. The one having only C atoms at both edges is defined as AC-CC/CC and the one with both C and N atoms at the edges are called as AC-CN/CN. The third type is the AC-C$_3$NNRs with combined edge structures of CC and CN, which means one edge is composed of only C atoms and the other edge is composed of both C and N atoms named as AC-CC/CN. Similarly, ZZ-C$_3$NNRs have three structures of ZZ-CC/CC, ZZ-CC/CN, ZZ-CN/CN, as shown in **Fig. 1(b)**.

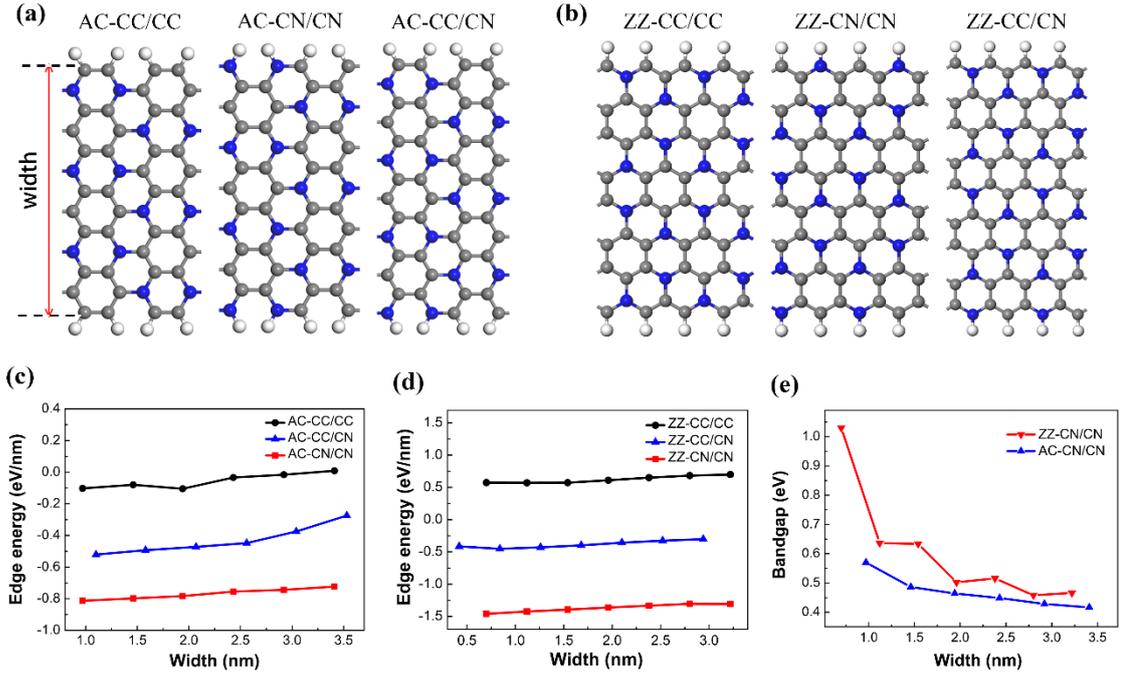

**Fig. 1.** (a-b) The atomistic configurations of (a) AC-C$_3$NNRs and (b) ZZ-C$_3$NNRs with three types of edge configurations. The grey, blue and white balls represent C, N and H atoms, respectively. (c-d) The edge energies of (c) AC-C$_3$NNRs and (d) ZZ-C$_3$NNRs as a function of the ribbon width. (e) The bandgaps of ZZ-CN/CN (red line) and AC-CN/CN (blue line) as a function of the ribbon width.

To analyze the thermodynamic stability of C$_3$NNRs with different edges, we calculated the edge energy ($E_{edge}$) by using the following equation [37,38],

$$E_{edge} = \frac{E_{ribbon} - n_C E_C - n_N E_N - n_H E_H}{2l} \quad (1)$$

where $E_{ribbon}$ is the total energy of the nanoribbon and $l$ represents the length of the ribbons. $E_C$, $E_N$ and $E_H$ are the energies of single C, N and H atoms in graphene, nitrogen and hydrogen molecules, respectively. $n_C$, $n_N$ and $n_H$ are the total numbers of C, N and H atoms in C$_3$NNRs, respectively. The effect of magnetism on the formation energy of different edges is also considered. The calculations demonstrated that the ground state of AC-C$_3$NNRs is nonmagnetic. As for ZZ-C$_3$NNRs, only ZZ-CN/CN is nonmagnetic, while both ZZ-CC/CC and ZZ-CC/CN are ferromagnetic. The calculated edge energies with respect to the ribbon width are shown in **Fig. 1(c-d)**. For both AC- and ZZ-C$_3$NNRs, it is obvious that CN/CN edge structure has the lowest edge energy, followed by CC/CN edge, and CC/CC edge has the lowest thermodynamic stability. $E_{edge}$ is also independent of the width of C$_3$NNRs. Moreover, it can be found that ZZ-CC/CC is energetically unfavorable due to its positive value of $E_{edge}$. ZZ-CN/CN has much lower value of $E_{edge}$ than AC-CN/CN at the same width, demonstrating that ZZ-CN/CN is more energetically favorable than AC-CN/CN.

The electronic properties of H-passivated C$_3$NNRs are calculated as well. The AC-C$_3$NNRs with three edge configurations are all semiconductors with bandgaps ranging from 0.7 eV to 0.4 eV within the width range of 0.5 nm to 4 nm (see **Fig. S2** in **SI**). Interestingly, although AC-CC/CN and AC-CC/CC have indirect bandgaps which is the same as that of C$_3$N nanosheet[11], the AC-CN/CN structure possesses the direct bandgap (see **Fig. S3-5** in **SI**). As for ZZ-C$_3$NNRs, only ZZ-CN/CN is a semiconductor with the indirect bandgap, while both ZZ-CC/CC and ZZ-CC/CN present metallic feature (see **Fig. S6-7** in **SI**). The bandgap of ZZ-CN/CN with the width of 0.7 nm is as large as 1.05 eV, while its bandgap decreases to 0.6 eV when the width of ZZ-CN/CN increases to 1.1 nm. With the continuing increase of the ribbon width, the bandgap of ZZ-CN/CN converges to a value of 0.45 eV up to the width of 3.2 nm.

**Fig. 1(e)** shows the predicted bandgaps of the most stable AC-CN/CN and ZZ-CN/CN with width varying from 0.7 to 3.5 nm. The bandgap of ZZ-CN/CN is larger than that of AC-CN/CN at the same width. Previous studies reported that the calculated (based on PBE functional) bandgap of C$_3$N sheet are 0.39 eV [11,13]. In general, the bandgaps of $C_3N$NRs is larger than that of the $C_3N$ sheet, especially when the nanoribbons are narrower than 2nm. This trend is consistent with previous work which revealed that cutting two-dimensional material into nanoribbons can enlarge the bandgap of the material[18]. It is well known that PBE functional always underestimates the bandgap, hence we also used HSE06 functional to obtain the more accurate bandgaps of nanoribbons. Due to the limitation of the computational model, only the bandgaps of several structures are calculated by HSE06 functional. The bandgaps calculated by PBE and HSE06 functionals are shown in **Table 1**. We can see that the bandgaps calculated by HSE06 functional are about 0.9eV larger than those calculated by PBE one.

**Table 1.** Comparison of the bandgaps calculated by PBE and HSE06 functionals. $\Delta E_b$ denotes that the bandgap calculated by HSE06 subtract that calculated by PBE.

|  | PBE (eV) | HSE06 (eV) | $\Delta E_b$ (eV) |
| --- | --- | --- | --- |
| AC-CC/CC (W=2.4 nm) | 0.44 | 1.31 | 0.87 |
| AC-CN/CN (W=2.4 nm) | 0.45 | 1.35 | 0.90 |
| AC-CC/CN (W=2.6 nm) | 0.46 | 1.36 | 0.90 |
| ZZ-CN/CN (W=1.1 nm) | 0.64 | 1.54 | 0.90 |

Based on the above calculations, we can see the bandgaps of $C_3N$NRs don't change too much with the width and edge structures when the width of ribbon is larger than 2 nm. Previous studies have shown that vdW stacking is another efficient way to tune the bandgap of 2D materials [39,40]. Therefore, we further evaluate the bandgaps of bilayer $C_3N$NRs. Since our calculations have demonstrated that CN/CN is the most energetically favorable edge structure for both AC-$C_3N$NRs and ZZ-$C_3N$NRs, we only considered the stacking of AC-CN/CN and ZZ-CN/CN in the following calculations. Four stacking structures of bilayer $C_3N$NRs, namely AA-, AA'-, AB- and AB'-stacking, are calculated. In the AA-stacking, all atoms in the lower layer are aligned with the upper layer and CC and NN overlaps are included, as shown in **Fig. 2(a-b)**. Analogously, in the AA'-stacking, all atoms in the lower layer are aligned with the upper layer but CC and CN overlaps are included. In the AB-stacking, half atoms of the lower layer are aligned with the center of the upper hexagon but the other half atoms are aligned with the upper atoms, in which both CC and NN overlaps are included. In the AB'-stacking, one half atoms of lower layer are aligned with the center of the upper hexagon and the other half atoms are aligned with the upper atoms, in which CC and CN overlaps are included. Relative to the strong covalent coupling in the plane, the interactions between the neighboring layers are the weak vdW force which is dependent on the interlayer distance. The interlayer distance of AC-CN/CN is 3.6 Å for AA-stacking and 3.4 Å for AA'-, AB- and AB'- stacking structures. As for ZZ-CN/CN, the interlayer distance is 3.7 Å for AA-stacking and 3.4 Å for AA'-, AB- and AB'-stackings. The interlayer distance of 3.4 Å is close to the interlayer distance of graphene which is 3.35 Å[41].

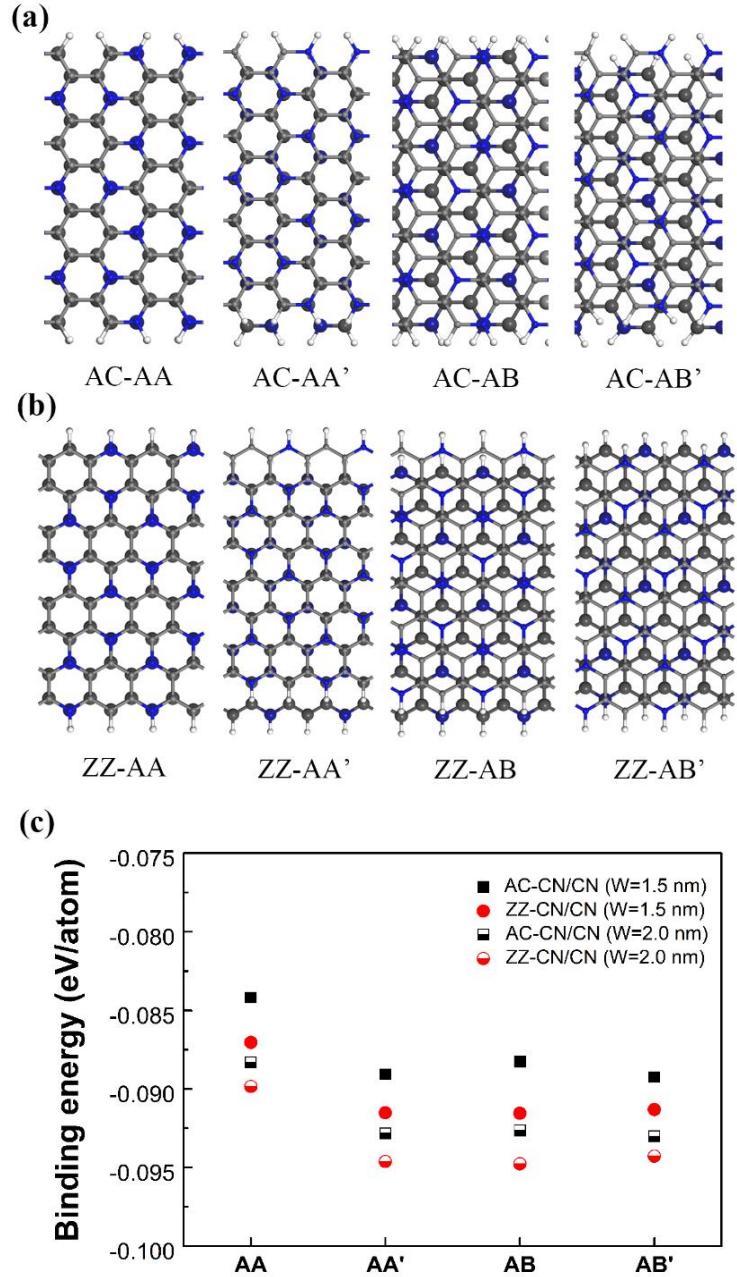

**Fig. 2** (a-b) Structure illustration of (a) bilayer AC-CN/CN and (b) bilayer ZZ-CN/CN with AA-, AA'-, AB- or AB'-stacking. The grey, blue and white balls represent C, N and H atoms, respectively. Besides, the lower atoms have deeper color and larger size than upper atoms. (c) Binding energies of four stacking orders with different edge configurations and widths. The black solid square and black hollow square stand for AC-CN/CN with the width of 1.5nm and 2.0nm, respectively. And the red solid circle and red hollow circle represent ZZ-CN/CN with the width of 1.5nm and 2.0nm, respectively.

To compare the thermodynamic stability of these stacking structures, the interlayer binding energy was calculated by the following formula,

$$E_{bind} = \frac{E_{bilayer} - 2 \times E_{monolayer}}{n} \quad (2)$$

where $E_{bilayer}$ and $E_{monolayer}$ are the energies of bilayer and monolayer nanoribbons, respectively. $n$ is the total number of atoms of the bilayer nanoribbon. **Fig. 2(c)** shows the binding energies of stacking structures of bilayer AC-$C_3$NNRs and ZZ-$C_3$NNRs with width of 1.5nm and 2.0nm. The binding energy decreases with the increase of ribbon width, and the binding energy of ZZ-CN/CN bilayer is lower than that of the AC-CN/CN one with the same width. Among different stackings, AB-, AB'- and AA'-stacking structures have similar levels of binding energies. AA-stacking structure has less interlayer binding because it has the most number of CC and NN overlaps which cause the strong repulsion between the two layers.

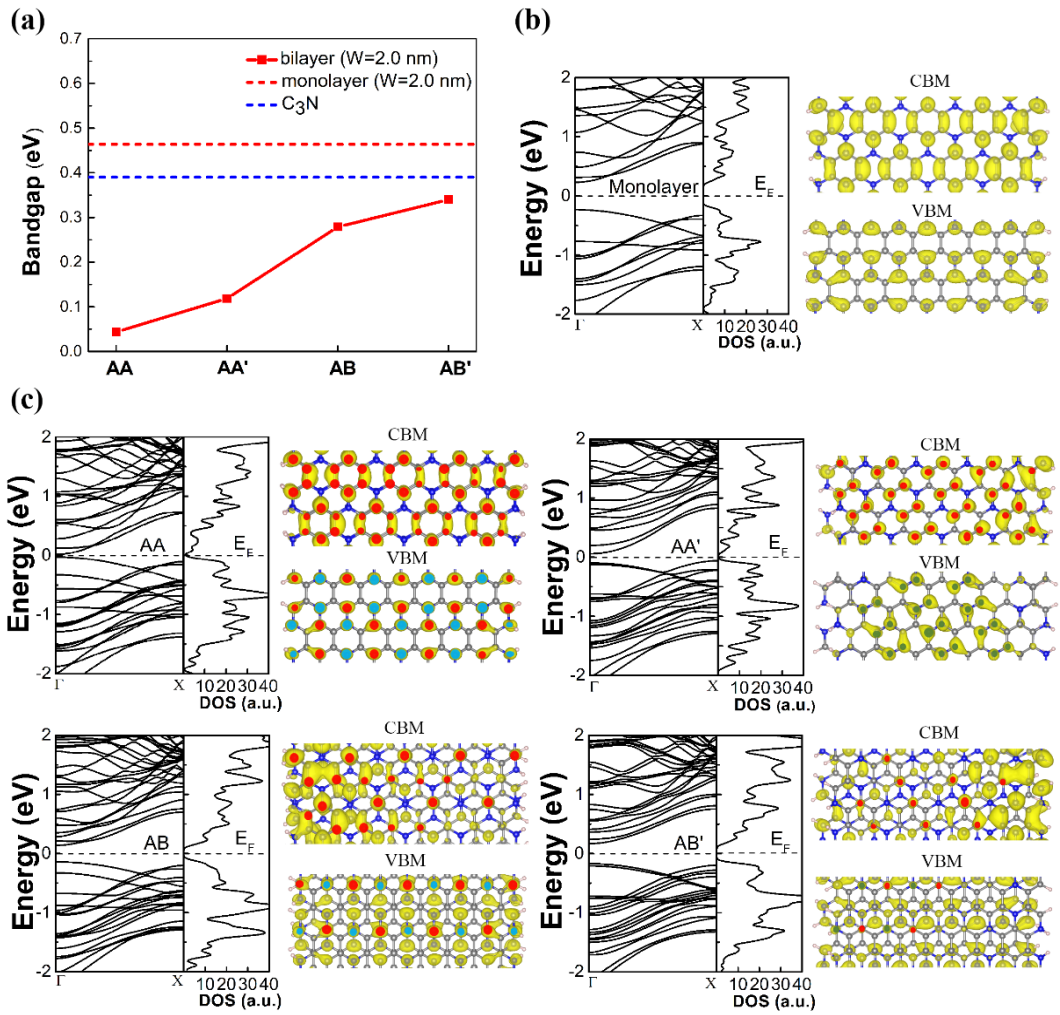

**Fig. 3** (a) The bandgaps of bilayer AC-CN/CN with AA-, AA'-, AB- and AB'-stacking, which is shown by red solid line. The red and blue dashed lines exhibit the bandgaps of monolayer AC-CN/CN and $C_3N$ sheet, respectively. (b)The band structure, density of states (DOS) and partial charge (CBM and VBM) of monolayer AC-CN/CN. (c)The band structure, DOS and partial charge (CBM and VBM) of bilayer AC-CN/CN with AA-, AA'-, AB- and AB'-stacking. The red, blue and green circles stand for $p_z$ orbital overlap contributed by CC, NN and CN atoms.

To explore the electronic properties of these heterostructures, the bandgaps of bilayer $C_3NNRs$ with different stacking orders were calculated. Here, only bilayer $C_3NNRs$ with CN/CN edge are considered in our calculations because of the energetical stability of this edge, as shown in **Fig. 1(c-d)**, and all the considered $C_3NNRs$ have a width of 2 nm. **Fig. 3(a)** shows the significant bandgap reduction of all bilayer AC-CN/CN with different stacking orders, in comparison with those of monolayer AC-CN/CN (red dash line) and $C_3N$ nanosheet (blue dash line). Specifically, compared with the monolayer AC-CN/CN, the bandgaps of bilayer AC-CN/CN heterostructures with AA-, AA'-, AB- and AB'-stacking reduce by 0.41eV, 0.34eV, 0.18eV and 0.12eV, respectively. These make the bandgaps of the four stacking structures follow the order of AB'- > AB- > AA'- > AA-stacking. To understand the effect of stacking order on the bandgaps of $C_3$NNRs, we calculated the band structures and density of states (DOS) of bilayer AC-CN/CN with different stacking orders, as well as their monolayer counterparts. Due to the vdW interlayer coupling, remarkable changes of the CBM and VBM are observed in bilayer heterostructures in comparison with those of the monolayer AC-CN/CN. As shown in **Fig. 3(c)**, bilayer AC-CN/CN have obvious energy downward shift of CBM and upward shift of VBM. The specific difference of the band structure between bilayer AC-CN/CN and monolayer AC-CN/CN can be seen from **Fig. S8(a)** in **SI**. The blue arrow represents energy shift, and the specific value can be seen from **Table 2**

To further understand the mechanism of energy shift induced by the stacking, we calculated the partial charge distributions of CBM and VBM. As mentioned above, CBM has a downward movement, and VBM has an upward movement. We ascribed such shift trend of CBM and VBM to the interlayer coupling, the strength of which depends on the number of orbital overlaps. For the AA-stacking in **Fig. 3(c)**, the partial charge distribution of CBM and VBM of upper and down layer are similar to that of monolayer. As a result, there are a lot of $p_z$ orbital overlap between layers, which are represented by red and blue circles in **Fig. 3(c)** (The red, blue and green circles represent $p_z$ orbital overlap attributed to CC, NN and CN atoms, respectively). As shown in **Table 2**, the AA-stacking has 50 orbital overlaps attributed to CC atoms overlap in CBM, and 18 CC and 18 NN overlaps in VBM. These orbital overlaps and interlayer coupling lead to the energy shift of CBM and VBM near the Fermi level, which finally result into significant reduction of bandgaps. As for the AA'-stacking, there are 32 CC overlaps in CBM and 22 CN overlaps in VBM due to the localized orbital distribution as shown in **Fig. 3(c)**.

For AB- and AB'-stacking, their orbital overlaps are much less than those of AA- and AA'-stacking, since there are half atoms of lower layer are aligned with the upper layer atoms. This indicates that the possibility of orbital overlap reduces by half at the least. Specifically, there exists 20 CC orbital overlaps in CBM of the AB-stacking, while there are 9 CC and 9 NN orbital overlaps in VBM of AB-stacking. It is clear that the number of orbital overlap is roughly reduced by half compared to the AA-stacking. Therefore, the energy shift of CBM and VBM in AB-stacking is significantly reduced compared with the AA-stacking. As for the AB'-stacking, there exists 10 CC orbital overlaps in CBM, while there are 4 CC and 4 NN orbital overlaps in VBM. Thus, the energy shift of the AB'-stacking is least among all stacking types. In addition to the bandgap change, the band structures of bilayer heterostructures also changes with the stacking order. For example, AA-, AB- and AB'-stacking structures keep the direct bandgap, while AA'-stacking has the indirect bandgap. Such a change of band structure feature is caused by the localized orbital distribution in AA'-stacking. As shown in **Fig. 3(c)**, compared with monolayer AC-CN/CN, the partial charge distribution of VBM in bilayer AC-CN/CN with AA'-stacking is much more localized.

**Table 2.** Orbital overlap and energy shift ($\Delta E$) of CBM and VBM in bilayer AC-CN/CN heterostructures.

|      | CBM   | $\Delta E$ (eV) | VBM         | $\Delta E$ (eV) |
|------|-------|-----------------|-------------|-----------------|
| AA   | 50 CC | 0.22            | 18 CC+ 18 NN | 0.21           |
| AA'  | 32 CC | 0.17            | 22 CN       | 0.17            |
| AB   | 20 CC | 0.09            | 9 CC+9 NN   | 0.09            |
| AB'  | 10 CC | 0.06            | 4 CC+4 CN   | 0.06            |

For bilayer ZZ-CN/CN heterostructures with AA-, AA'-, AB- and AB'-stacking, the bandgaps reduced by 0.32eV, 0.26eV, 0.13eV and 0.05eV, respectively. This also makes the bandgaps of the four stacking structures follow the order of AB'- > AB- > AA'- > AA-stacking, which is similar with the case of bilayer AC-CN/CN heterostructures. As shown in **Fig. 4 (a)**, all of the bilayer ZZ-CN/CN heterostructures show bandgap reduction in comparison with monolayer ZZ-CN/CN (red dash line). While the bandgap of AB'-stacking is smaller than that of monolayer ZZ-CN/CN (dash red line) but larger than that of $C_3N$ nanosheet (blue dash line). Similar to bilayer AC-CN/CN heterostructures, the energy shift of CBM and VBM in bilayer ZZ-CN/CN is also dependent on the number of orbital overlaps. As shown in **Fig. 4(c)** and **Table 3**, the number of orbital overlaps follows an order of AA- >AA'- >AB- > AB'-stacking, so the energy shift of AA-stacking is the largest while AB'-stacking is the least. The specific difference of the band structure between bilayer ZZ-CN/CN and monolayer ZZ-CN/CN can be seen from **Fig. S8(b)** in **SI**. More interestingly, as shown in **Fig. 4(c)**, AA-, AA'- and AB'-stacking keep the feature of indirect bandgap, while AB-stacking transforms into direct bandgap structure. From the partial charge distributions of CBM and VBM in AB-stacking structure, we can see the obvious localization of partial charge distribution in this structure. Such a localization causes the band structure transition from indirect to direct one. We also used HSE06 functional to check our results. Due to the bilayer system is too large to calculate by HSE06, we only checked the band structure of monolayer. It can be seen that the band structure calculated by HSE06 functional is the same as that calculated by PBE. The only difference is that the bandgap calculated by HSE06 functional is larger than those calculated by PBE one. (see **Fig. S9** in **SI**)

The bandgap tuning of $C_3NNRs$ shown in our study provides $C_3N$ with wider range of bandgap which is crucial to $C_3N$'s applications in electronics. Especially, the transformation of indirect to direct bandgap endows $C_3NNRs$ with promising applications in optoelectronic devices and photocatalysis.

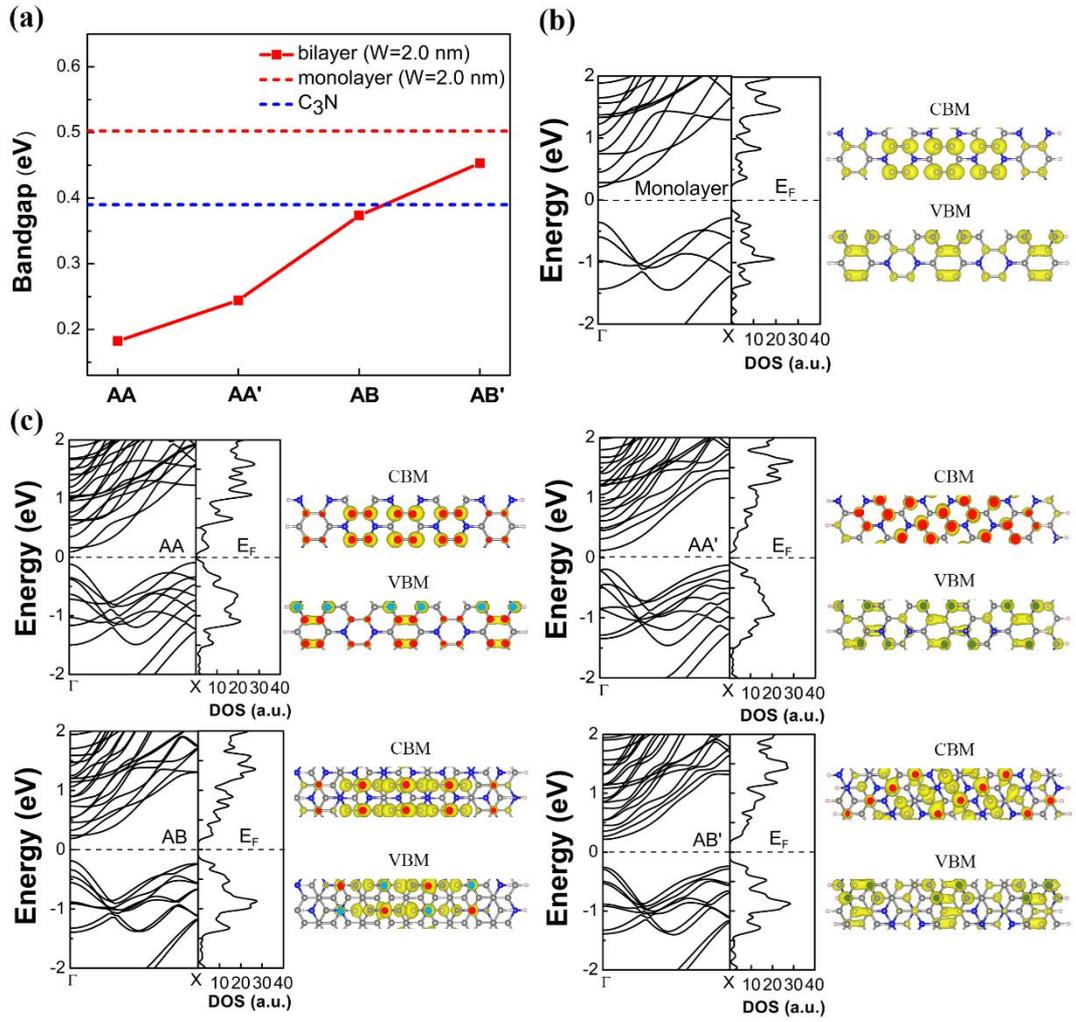

**Fig. 4.** (a) The bandgaps of bilayer ZZ-CN/CN with four stacking orders, which is shown by red solid line. The red and blue dashed line show the bandgaps of monolayer ZZ-CN/CN and $C_3N$ sheet. (b-c) The band structure, DOS and partial charge (CBM and VBM) of (b) monolayer ZZ-CN/CN and (c) bilayer ZZ-CN/CN heterostructure with different stacking orders. The red, blue and green circles represent $p_z$ orbital overlap of CC, NN and CN atoms, respectively.

**Table 3.** Orbital overlap and energy shift ($\Delta E$) of CBM and VBM in bilayer ZZ-CN/CN heterostructures.

|       | CBM    | $\Delta E$ (eV) | VBM        | $\Delta E$ (eV) |
|-------|--------|-----------------|------------|-----------------|
| AA    | 20 CC  | 0.12            | 20 CC+6 NN | 0.20            |
| AA'   | 18 CC  | 0.09            | 10 CN      | 0.17            |
| AB    | 10 CC  | 0.03            | 4 CC+4 NN  | 0.10            |
| AB'   | 10 CC  | 0.00            | 6 CN       | 0.05            |

## 4. Conclusions

In summary, we carried out systematic investigations on the atomistic structures, energetic stabilities and electronic properties of monolayer and bilayer $C_3$NNRs by first-principles calculations. Our computational results indicated that cutting $C_3$N into nanoribbons is an effective way to enlarge its bandgap and the bandgaps of $C_3$NNRs exhibit monotonic reduction with the increase of their widths. Moreover, AC-CN/CN nanoribbons are direct bandgap semiconductors, while AC-CC/CC, AC-CC/CN and ZZ-CN/CN are indirect bandgap semiconductors. Interestingly, the electronic properties of monolayer $C_3$NNRs can be significantly tuned by bilayer stacking. For both AC- and ZZ-$C_3$NNRs, bilayers with AA- or AA'-stacking have smaller bandgaps than those with AB- or AB'-stacking. The larger bandgap reduction of AA- or AA'-stacking is attributed to the more orbital overlap and thus more band shift near the Fermi level. Particularly, bilayer AC-CN/CN with AA-, AB- and AB'-stacking remain as direct bandgap semiconductor while AA'-stacking becomes indirect bandgap semiconductor because of the localization of charge distribution. As for bilayer ZZ-CN/CN, all the stacking structures keep the indirect bandgap feature but the bilayer with AB-stacking transforms into direct bandgap. This study predicts the effective engineering of $C_3$NNRs' electronic properties through stacking, and the tuning of bandgap and indirect-to-direct bandgap transformation endows $C_3$N with potential applications in electronic and optoelectronic devices.

**Acknowledgements**

Jia Liu and Xian Liao contributed equally to this work. This work was supported by the National Natural Science Foundation of China (Grant No. 21673075). The computations were performed in the ECNU Multifunctional Platform for Innovation (001).

# Supplementary Information

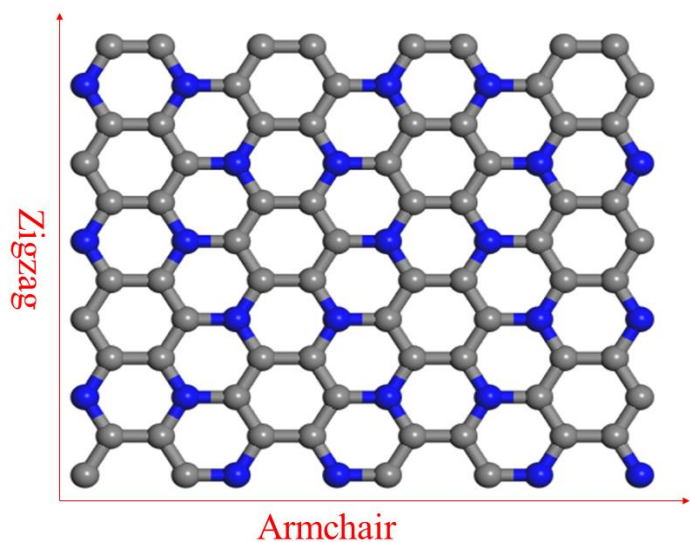

**Fig. S1** The top view of $C_3N$ sheet. The grey and blue balls represent C and N, respectively.

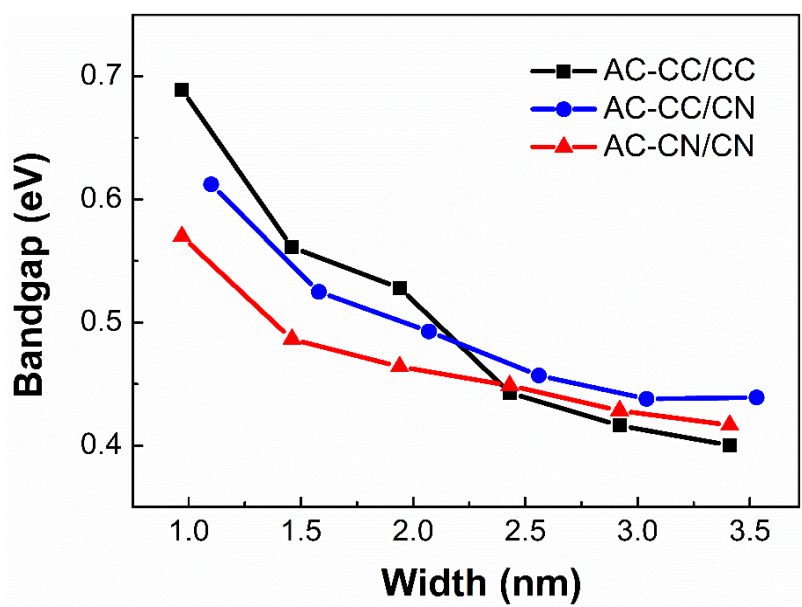

**Fig. S2** The bandgaps of AC-CC/CC (black line), AC-CC/CN (blue line) and AC-CN/CN (red line) with the increase of nanoribbon width.

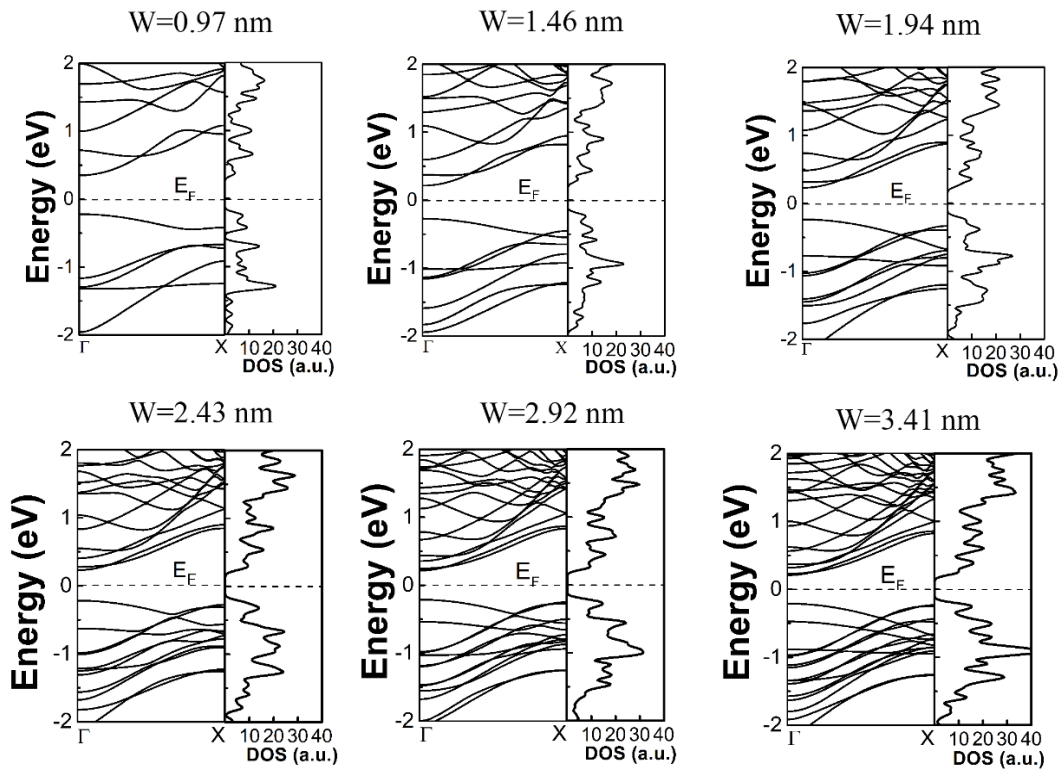

**Fig. S3** The band structure and DOS of monolayer AC-CN/CN nanoribbons with different widths.

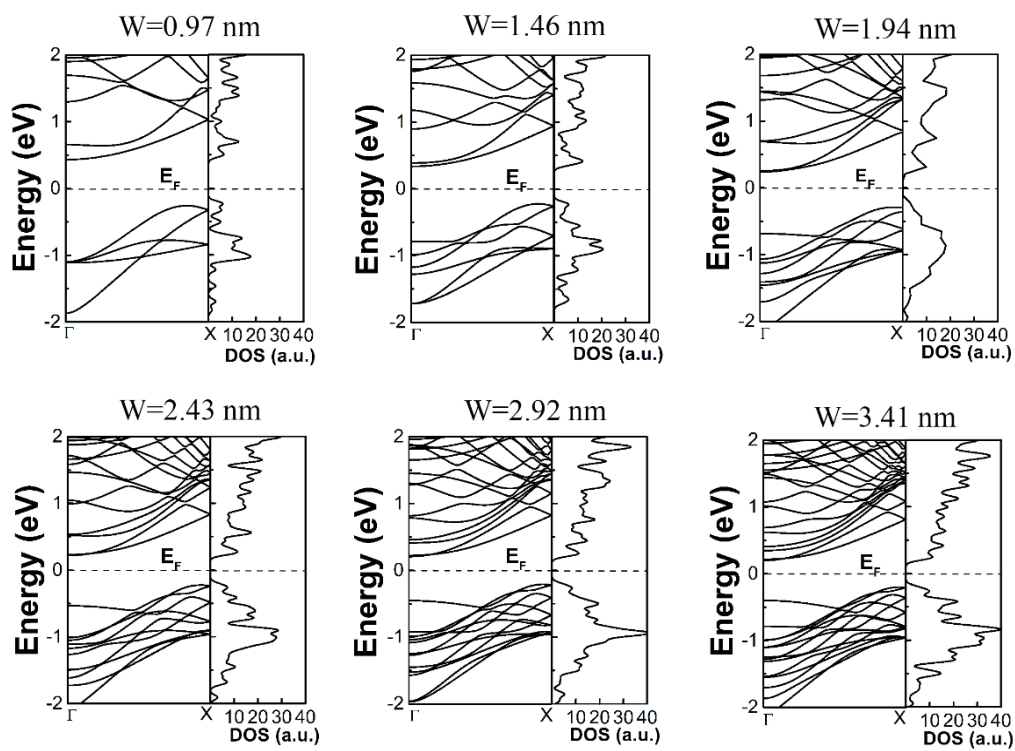

**Fig. S4** The band structure and DOS of monolayer AC-CC/CC nanoribbons with different widths.

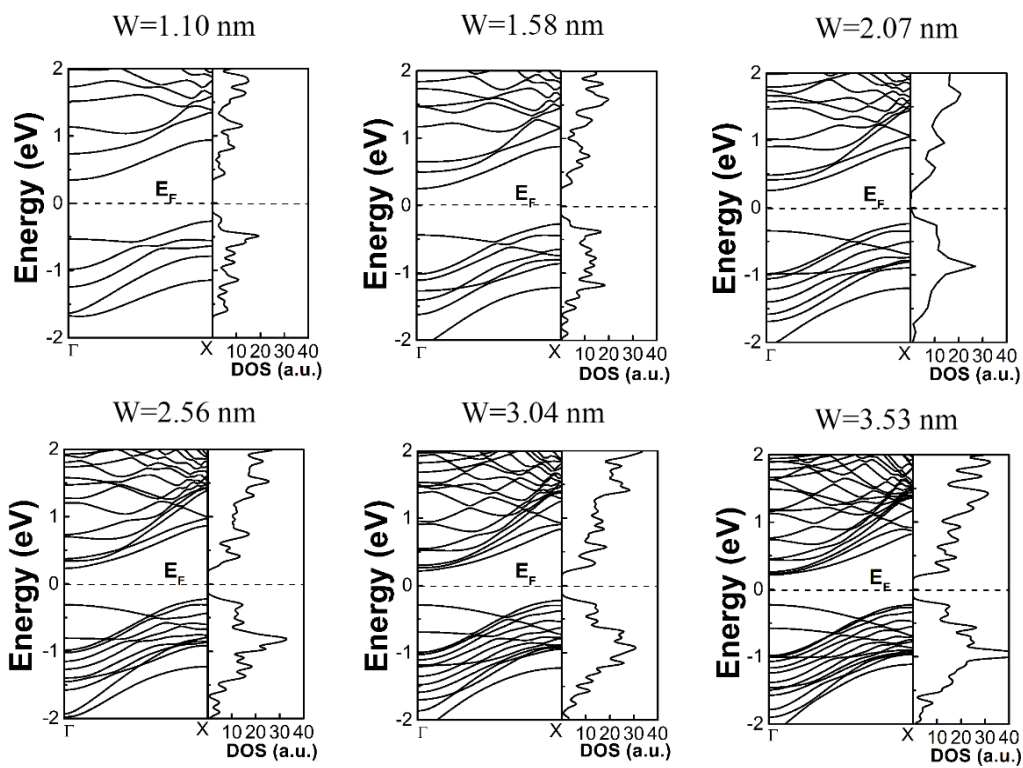

**Fig. S5.** The band structure and DOS of monolayer AC-CC/CN nanoribbons with different widths.

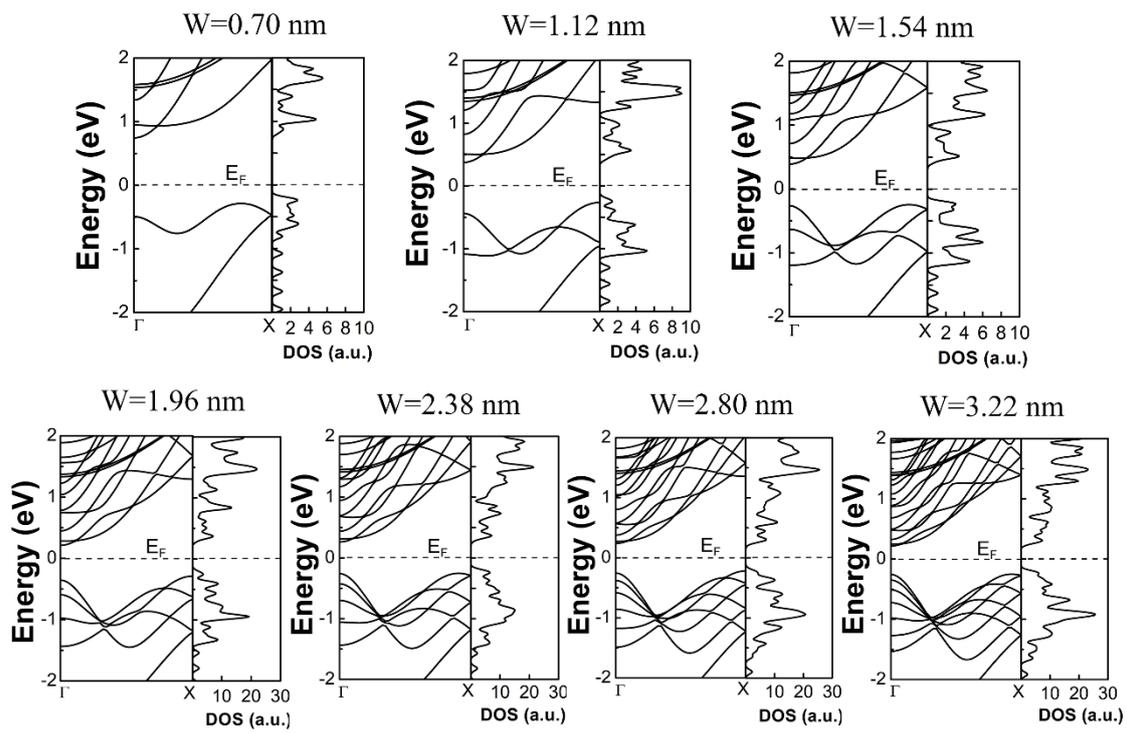

**Fig. S6** The band structure and DOS of monolayer ZZ-CN/CN nanoribbons with different widths.

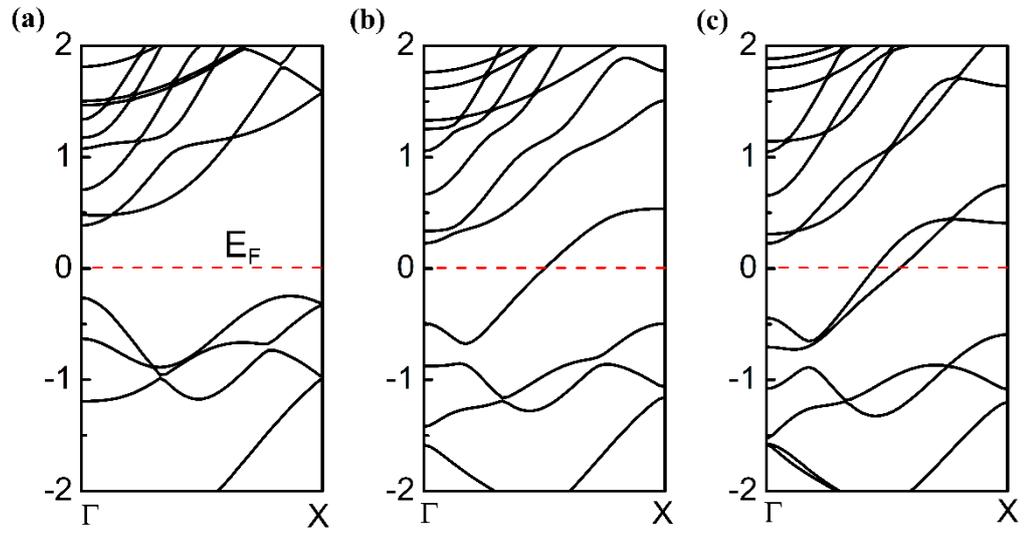

**Fig. S7** The band structure of monolayer (a) ZZ-CN/CN with the width of 1.5 nm, (b) ZZ-CC/CN with the width of 1.3 nm and (c) ZZ-CC/CC with the width of 1.5 nm. The red dashed lines display the position of the Fermi level.

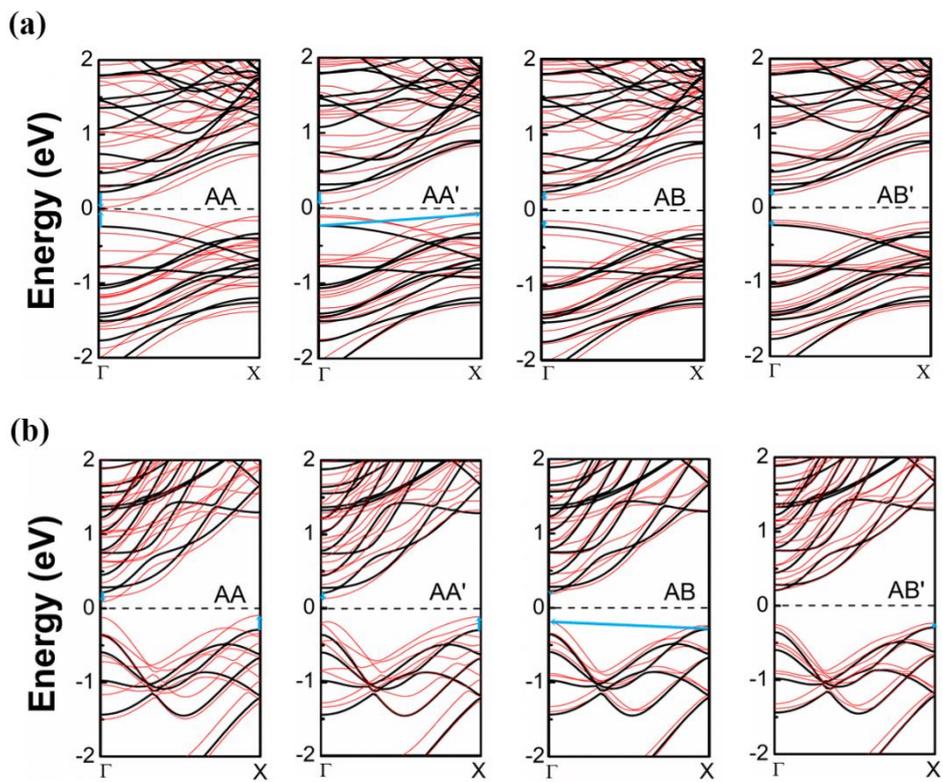

**Fig. S8** (a) The difference of band structure between bilayer and monolayer AC-CN/CN with the width of 2 nm. (b) The difference of band structure between bilayer and monolayer ZZ-CN/CN with the width of 2 nm. The band structure of the monolayer is black, and the bilayer is red. The blue arrow represents the energy shift.

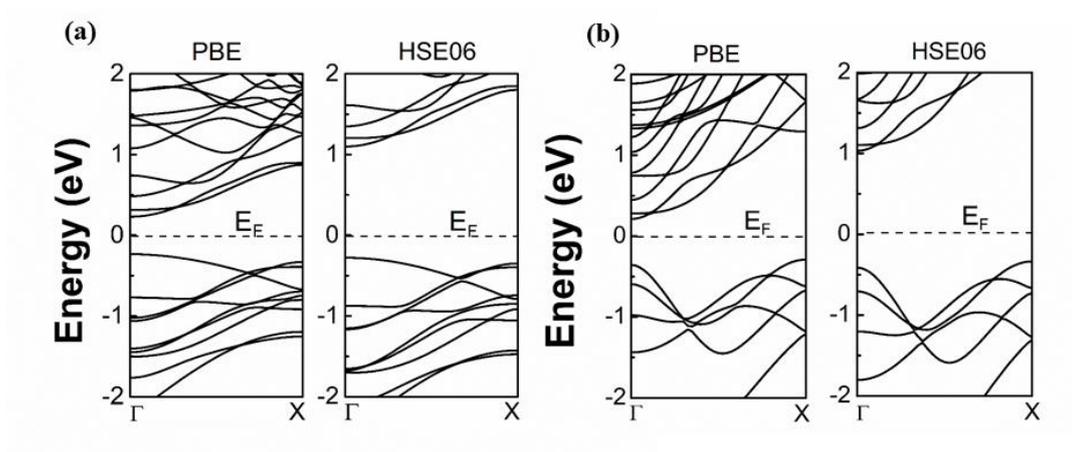

**Fig. S9** The band structure of monolayer (a) AC-CN/CN and (b) ZZ-CN/CN with the width of 2 nm by PBE and HSE06 functional.